\documentclass[a4paper,fleqn,usenatbib]{mnras}

\usepackage{newtxtext,newtxmath}

\usepackage[T1]{fontenc}
\usepackage{ae,aecompl}

\usepackage{graphicx}	
\usepackage{amsmath}	
\usepackage{amssymb}	

\def\ga{\mathrel{\hbox{\rlap{\hbox{\lower4pt\hbox{$\sim$}}}\hbox{$>$}}}}
\def\la{\mathrel{\hbox{\rlap{\hbox{\lower4pt\hbox{$\sim$}}}\hbox{$<$}}}}

\def\lmin{$L_{\rm min}$}

\def\lbol{$L$\mbox{$_{\rm bol}$}}
\def\lmin{$L_{\rm min}$}

\def\teff{$T_{\rm eff}$}

\def\logl{$\log (L/L_\odot)$}

\def\AV{$A_{V}$}

\title[Discrepancies in the ages of young star clusters]{Discrepancies in the ages of young star clusters; evidence for mergers?}

\author[E. R. Beasor et al.]{
Emma R. Beasor$^{1}$\thanks{E-mail: e.beasor@2010.ljmu.ac.uk},
Ben Davies$^{1}$,
Nathan Smith$^{2}$
and Nate Bastian$^{1}$
\\
$^{1}$Astrophysics Research Institute, Liverpool John Moores University, 146 Brownlow Hill, Liverpool L3 5RF, UK\\
$^{2}$Steward Observatory, University of Arizona, 933 N. Cherry Ave., Tucson, AZ 85721, USA
}
\date{Accepted XXX. Received YYY; in original form ZZZ}

\pubyear{2015}

\begin{document}
\label{firstpage}
\pagerange{\pageref{firstpage}--\pageref{lastpage}}
\maketitle

\begin{abstract}
There is growing evidence that star clusters can no longer be considered simple stellar populations (SSPs). Intermediate and old age clusters are often found to have extended main sequence turn-offs (eMSTOs) which are difficult to explain with single age isochrones, an effect attributed to rotation. In this paper, we provide the first characterisation of this effect in young (<20Myr) clusters. We determine ages for 4 young massive clusters (2 LMC, 2 Galactic) by three different methods: using the brightest single turn-off (TO) star; using the luminosity function (LF) of the TO; and by using the lowest \lbol\ red supergiant (RSG). The age found using the cluster TO is consistently younger than the age found using the lowest RSG \lbol. Under the assumption that the lowest luminosity RSG age is the `true' age, we argue that the eMSTOs of these clusters cannot be explained solely by rotation or unresolved binaries. We speculate that the most luminous stars above the TO are massive blue straggler stars formed via binary interaction, either as mass gainers or merger products. Therefore, using the cluster TO method to infer ages and initial masses of post-main sequence stars such as Wolf-Rayet stars, luminous blue variables and RSGs, will result in ages inferred being too young and masses too high.
\end{abstract}

\begin{keywords}
stars: evolution -- galaxies: clusters -- stars: massive -- blue stragglers
\end{keywords}



\section{Introduction}
Historically, star clusters have been used as a benchmark for stellar evolution models, since it was assumed they were simple stellar populations (SSPs) of a single age and metallicity. However, modern observations of intermediate age and old globular clusters reveal features not easily explained by SSPs, for example the presence of multiple main sequences \cite[e.g.][]{marino2008spectroscopic,piotto2015hubble}, extended main sequence (MS) turnoffs \citep[e.g.][]{keller2011extended, bastian2016young} and abundance anticorrelations \cite[e.g.][]{gratton2004ac, gratton2012ac}. Within the literature there are a number of explanations for these peculiar features. Recent work suggests that the extended MS turnoff is an effect of stellar rotation, not age spreads as previously thought \citep[see][for a recent review]{bastianmultipops}. This can be understood as rotational mixing lengthens the MS lifetime by injecting more hydrogen into the core, also causing the terminal age MS stars to appear more luminous.   

So far, extended main sequence turn-offs (eMSTOs) are found to exist in intermediate and old age clusters. If the eMSTO phenomenon is caused by stellar rotation, then young (<50Myr) clusters should also exhibit this feature, since rotation is thought to have a greater influence the evolution of more massive stars. Indeed, eMSTOs have been observed in clusters as young as 20-30Myrs \citep{li2018discovery}.

One other potential explanation for stars above the MSTO is rejuvenation, either via mergers or mass transfer \cite[e.g.][]{schneider2014ages,gosnell2015implications}. It is now commonly believed that most massive stars exist in binaries \citep[$\geq$ 70\%, ][]{sana2012binary} and the merger rate of these could be high \citep{demink2014incidence}. For stars undergoing binary interaction, mass can be transferred from a secondary on to a primary (mass-gainer) or the two objects can merge entirely. Not only does the merger product gain mass and hence become more luminous, but its MS lifetime is extended by additional hydrogen being mixed into the core. Merger products would therefore appear as younger, brighter stars above the TO, much like blue straggler (BS) stars seen in globular clusters \citep{sandage1984brightest,knigge2009binary, gosnell2015implications}. BSs have been tentatively observed in the young cluster NGC 330 \citep{lennon1993spectroscopic,evans2006vlt} and in Westerlund 1 \citep{clark2018vlt}, but there has been no systematic study on whether they routinely exist in young clusters. If ubiquitous, the eMSTOs in young clusters would impact their age estimates in the literature.   

In previous papers we have determined the ages to young clusters using two different methods. The first, for NGC 2100, involved fitting the full observed luminosity range of the red supergiants \citep[RSGs, ][] {beasor2016evolution} with predictions from various stellar evolution models. In \citet{beasor2018evolution} we used the previously determined ages for $\chi$ Per and NGC 7419, in both cases found by fitting isochrones to MS stars by eye \citep[14 Myr for both, ][]{currie2010stellar,marco2013ngc}. For both of these methods we assumed only single stellar evolution (no binaries, no mergers). 

Each age estimation technique has its potential weaknesses. If the merger rate in young clusters is high \citep[as suggested by][]{demink2014incidence}, post-merger objects (i.e. an object resulting from the merging of two stars in a binary system) could affect the age determined in each method significantly, as they would appear as bright objects above the TO.  Once these stars leave the MS, they could evolve to become RSGs which are anomalously bright compared to the single star population, again causing the observer to infer a younger age for the cluster. 
  
In this paper we compare ages for 4 young clusters using three different methods and attempt to reconcile these differences in terms of non-simple stellar evolution. In Section 3 we describe each of the three age determination methods in detail as well as the method for determining extinction towards the cluster. In Section 4 we discuss the results of this work and finally in Section 5 we discuss the implications of our results with respect to possible evidence for mergers.

\section{Observations}
\subsection{Sample}
For this work we require clusters for which we are able to obtain age estimations from the MS turnoff {\it and} the RSGs. The clusters must therefore be spatially resolvable, and be young enough and massive enough to have a well-populated RSG branch. Based on this, we have chosen 2 young clusters from within the Galaxy (NGC 7419, $\chi$ Per) and 2 located in the LMC (NGC 2100, NGC 2004). 

\subsection{Photometry}
The majority of the data used in this work is archival. We will now briefly describe the MS and RSG data for each cluster. The MS data for NGC 7419 is $UBV$ photometry from \citet{beauchamp1994galactic}. The RSG photometry includes $UBVRI$ from \citet{joshi2008multiwavelength}, and near and mid-infrared (IR) photometry from 2MASS \citep{skrutskie2006two} and MSX \citep{price2001midcourse}. 

The main sequence photometry for $\chi$ Per is from \citet{currie2010stellar}. As this catalogue contained data for both clusters in the $h$ + $\chi$ Per complex, we simply included any star that was within 6 arcmin of the cluster centre, the distance to the edge of the complex. For the RSGs, we used archival photometry from \citet{johnson1966law, kharchenko2009integrated,pickles2010all} and MSX \citep{price2001midcourse}. As described in \citet{davies2018initial}, there was an offset in the $I$-band between the Johnson and Pickles photometry which could not be explained. We have again taken an average of both the measurements and defined the error to be half the difference between the two.   

For NGC 2004 and NGC 2100, we use the dereddened photometry from \citet{niederhofer2015no} for the MS stars, originally from \citet{brocato2001large}. This photometry has been dereddened to mitigate the effect of differential extinction using the method described in \citet{milone2012acs}. The RSG photometry is compiled from Spitzer, MSX and WISE \citep{werner2004spitzer,price2001midcourse,wright2010wide}. 
\subsection{Distances}

To determine the distances to the two Galactic clusters, we first isolate all hot star cluster members by searching the SIMBAD database for OB stars in the plane of each cluster. Next, we obtain Gaia DR2 parallaxes and proper motions for all stars. Following \citet{aghakhankoo2019inferring} the error on the parallax $\sigma_i$ of each star $i$ is defined to be $\sigma_i = \sqrt{(1.086\sigma_\omega + \epsilon_i)}$ where $\epsilon_i$ is the excess astrometric noise on the Gaia parallax solution. We then iteratively sigma-clip the sample by discarding those stars with proper motions outside 3$\sigma$ of the mean for the whole cluster. Of the $N$ remaining stars (140 for $\chi~Per$, 20 for NGC~7419), we determine the sigma-weighted mean of the parallaxes $\bar{\pi}$ and the standard deviation $\sigma_{\bar{\pi}}$, and define the formal error on $\bar{\pi}$ to be $d\bar{\pi} = \sigma_{\bar{\pi}} / \sqrt{(N)}$. 

The posterior probability distribution function $P_r$  on the distance $r$ to the cluster is determined from $P_r \propto \exp(-0.5z^2)$, where $z = (\bar{\pi} - \pi_{\rm ZP} - 1/r)/d\bar{\pi}$, and $\pi_{\rm ZP}$ is the zero-point parallax offset in Gaia DR2\footnote{Unlike several other studies of Gaia DR2 distances, we do not adopt any Bayesian prior.}. Various independent measurements of $\pi_{\rm ZP}$ have been made, with a global average seeming to converge on $\pi_{\rm ZP} = -0.05$mas \citep[see][ for a discussion]{aghakhankoo2019inferring}. However, \citet{lindegren2018gaia} find that this value varies across the sky by $\pm0.03$mas on spatial scales of about a degree (i.e.\ larger than the cluster field of view). We therefore adopt a value of $\pi_{\rm ZP} = -0.05\pm0.03$mas, where this error on the zero-point offset must be added in quadrature to that of $\bar{\pi}$. 

For $\chi$~Per and NGC~7419, we find average parallaxes of $0.448 \pm 0.003$mas and $0.333 \pm 0.009$mas respectively, not including the uncertainty on $\pi_{\rm ZP}$. These values are converted to distances and uncertainties by finding the mode and 68\% confidence intervals of $P_rd$, finding $2.25^{+0.18}_{-0.14}$kpc and $2.93^{+0.32}_{-0.26}$ kpc for $\chi$~Per and NGC~7419 respectively. For further discussion on the distances see the forthcoming paper by Davies \& Beasor (submitted).

As NGC 2100 and NGC 2004 are both LMC clusters and we take the distance of 50$\pm$0.1 kpc \citep{lmcdist}.
\newline

\begin{table*}
\centering

\caption{Distances and extinctions for the clusters. The distances to the LMC clusters are taken from \citet{lmcdist}.}
\label{results}
\begin{tabular}{lccccccccc}

\hline
Cluster & Distance (kpc) & Extinction (\AV) \\
\hline
NGC 7419 &$2.93^{+0.32}_{-0.26}$ & 6.33 $\pm$ 0.22 \\
$\chi$ Per & $2.25^{+0.18}_{-0.14}$& 1.22 $\pm$ 0.22 \\
NGC 2100 & 50$\pm$0.1 & Differential, taken from \citet{niederhofer2015no} \\
NGC 2004 & 50$\pm$0.1 & Differential, taken from \citet{niederhofer2015no} \\

\hline\hline

\end{tabular}
\end{table*}

\section{Age estimations}
We now describe how we derive the extinction values for each cluster, and detail each of the three independent age determination methods. As we are using the de-reddened photometry for NGC 2004 and NGC 2100 we assume the extinction on this data should be negligible \citep[data taken from][]{niederhofer2015no} . 

\subsection{Estimating the foreground extinction}\label{extinction}
To estimate foreground extinction we began by constructing a colour-magnitude diagram (CMD) for each cluster. To isolate the MS we make cuts in colour-magnitude space. For the magnitude cut, we cut any bright stars that sit more than 2 magnitudes from the main sequence. For the faint end of the MS, we cut any stars obviously fainter than the point at which the sample is no longer complete (i.e. the brightness at which the number of stars per magnitude bin starts to decrease). For the colour cut, we discount anything that is clearly too red to be a member of the MS. 

To determine the best fitting extinction we employ population synthesis analysis, using a grid of MIST isochrones from ages 2Myrs to 100Myrs \citep{paxton2010modules,paxton2013modules,paxton2015modules,dotter2016mesa} at the approximate metallicity (+0.0 dex for Galactic clusters, -0.5 dex for LMC clusters). We begin by generating a population of 10000 stars between masses of 1.5 and 100M$_\odot$ sampled from the Salpeter initial mass function (IMF). We then interpolate the random stars onto the MIST mass-brightness function for whichever filters are appropriate for the photometry of the cluster being analysed. We adjust the synthetic photometry for the distance estimated in Sect 2.3 and a dummy input extinction value, according to the extinction law of \citet{cardelli1988ext}. We make the same cuts in magnitude and colour as we have done for the observed data.

\begin{figure}
\caption{Hess diagram for NGC 7419. The black points indicate stars that were included in the analysis. The red dashed line shows the fainter magnitude cut. }
\centering
\label{fig:heatmap}
\includegraphics[width=\columnwidth]{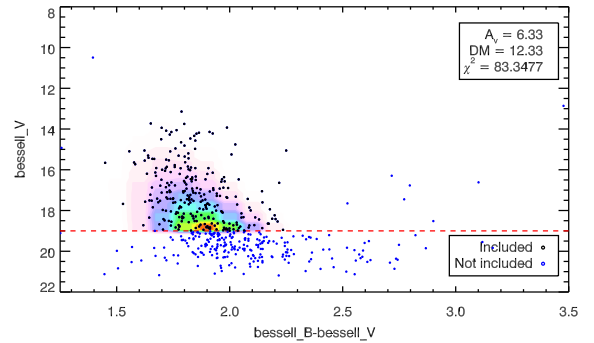}
\end{figure}

Next, we take both the synthetic MS and the observed MS and bin the stars into 2D histograms in colour-magnitude space. We then normalise the model distribution such that the model and observed CMDs have the same number of stars above the magnitude cut. We then compute the $\chi^2$ statistic between the observed and model histograms. This is repeated for a grid of input extinctions and ages, with the best fit extinction determined from interpolating where the $\chi^2$ goes to zero. Figure \ref{fig:heatmap} shows a Hess diagram for NGC 7419. Extinction estimates are shown in Table 1.

Note that, in principle, we can obtain an age estimate from this analysis. However, since the cluster age and extinction are virtually orthogonal to each other in terms of how they displace the MS on the CMD, we chose to simplify the analysis and first fix for extinction then do a single-parameter fit for the age. 

\subsection{Brightest turn-off star method}
The TO of a star cluster is defined as the most massive star that has yet to deplete H from its core. Since stellar mass and MS lifetime are strongly correlated, this can give an indication as to the cluster age.
Once we have identified the MS via the method described in Section \ref{extinction}  we then assume the brightest single star in this subsample is the TO point for the cluster.  Next, we use a grid of MIST isochrones (ages ranging from 2Myrs to 100Myrs, repeated for both rotating and non-rotating models), corrected for distance and extinction using the values found in the previous section, and identify the magnitude of the TO at each age (we assume the end of the MS is the point at which the central hydrogen mass fraction is less than 1x10$^{-4}$). From this we identify the theoretical TO magnitude-age relation. We then take the magnitude of the brightest TO star in the cluster and interpolate this onto the magnitude-age relation and hence derive an age for the cluster. Errors come from the photometric error on the magnitude. Results are shown in Table 2. 

One obvious weakness of this method is the susceptibility to stochastic effects from sparsely sampling the IMF at the high mass end. We look to improve upon this in Section 3.3. In addition, the presence of an eMSTO or bright BS-like stars above the TO would cause a younger age to be inferred. 

\subsection{Luminosity function}
\begin{figure}
\caption{\textit{Top panel:} Best fitting luminosity function for NGC 7419. The black line shows the observed TO LF while the blue line shows the model. \textit{Bottom panel:} Plot showing the error estimation for the LF method. The solid green line shows the best fitting age and the dashed green lines show the $\chi^2$ acceptability limit, i.e. $\chi^2_{\rm min}$+2.3.}
\centering
\label{fig:LF}
\includegraphics[width=\columnwidth]{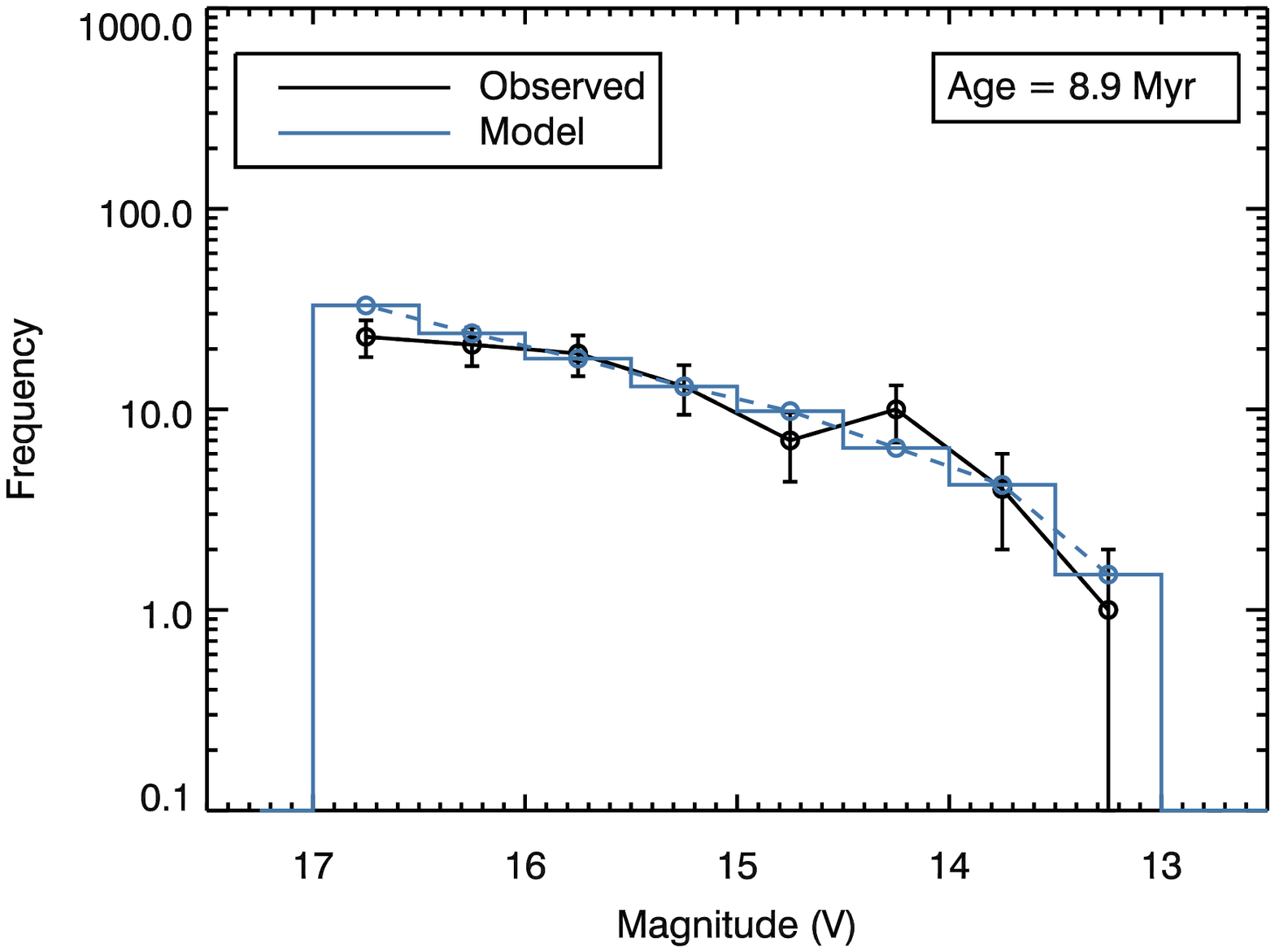}
\includegraphics[width=\columnwidth]{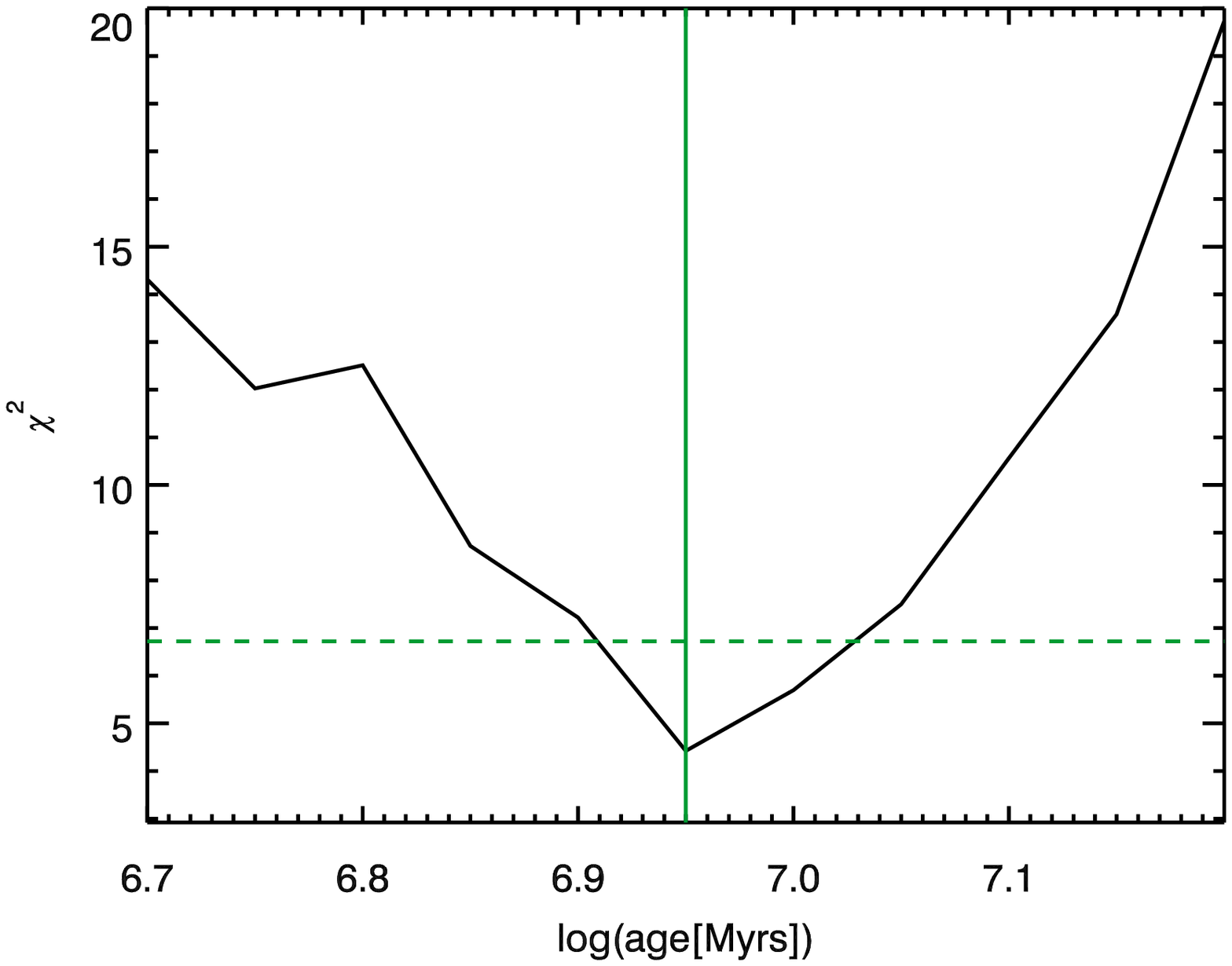}
\end{figure}

A similar but slightly more sophisticated method to determine the age from the turn-off is to model the luminosity function (LF) of the brightest stars in the main sequence. We use population synthesis (as described in Section \ref{extinction}) and adjust the colour and magnitude of the synthetic stars to the distance of the cluster and redden the photometry using the extinction derived in Section 3.1. For each age, histograms are then created for both the real and the synthetic stars of the number of stars per magnitude bin. These distributions are then compared and the best fit age is found using $\chi^2$ minimization (see below). Results are shown in Table 2 and an example best fit is shown for NGC 7419 in the top panel of Fig. \ref{fig:LF}. 

The best fitting age and 1-sigma error range is determined following \citet{anvi1976energy}. Our fitting function, derived from our population synthesis, has two degrees of freedom: age $t$ (which moves the LF left and right), and mass $M$ (which affects the normalization). We find the optimal age by minimizing the function,

\begin{equation}
\displaystyle
\chi^2 = \sum_i \frac{[ O_i - E_i(t,M) ]^2}{O_i}
\end{equation}

\noindent where $O_i$ and $E_i$ are the numbers of stars in the $i$th bin of the observed and model LFs respectively. The 68\% confidence intervals on $t$ are determined from those models with $\chi^2$ within 2.3 of the minimum, see bottom panel of Fig. \ref{fig:LF}.
   
The advantage of qualitatively fitting the LF at the TO, rather than simply taking the magnitude of the brightest turn-off star, is that it attempts to compensate for stochastic sampling of the TO. However it will still be affected by the presence of BSSs and stellar rotation. 
   
\subsection{Lowest luminosity red supergiant}
An independent method to determine the cluster age is to model the RSG population. Specifically, we use the bolometric luminosity (\lbol ) of the least luminous RSG in each cluster ($L_{\rm min}$). When a star crosses the HRD it does so at a nearly constant luminosity, and upon joining the RSG branch there is an initial decrease in \lbol. Once the star has settled on the RSG branch it returns to a luminosity similar to that when it left the MS. For an isolated RSG, it is hard to know the evolutionary history, i.e. how evolved it is and whether or not it descended from a single star or from a BS type object. However, in a large cluster of stars all born at the same time, the RSG with the lowest \lbol\ is most likely that which has evolved as a single star and most recently joined the RSG branch. 

To calculate \lbol\ we take all of the available photometry for a given star and integrate under the SED from blue to the mid-IR, as described in \citet{davies2018humphreys}. This method relies on the assumption that the flux emitted is spherically symmetric and as RSGs are known to have circumstellar extinction we also assume any flux lost at shorter ($<$U - band) wavelengths is reemitted at longer wavelengths. To account for any flux at shorter wavelengths we have extrapolated the SED using a blackbody temperature of 3500K, though in practice this region of the SED accounts for less than 0.01dex of the bolometric luminosity. 

Due to the steepness of the IMF  the number of RSGs in a given cluster is usually small, on the order of $\sim$10, leaving this method susceptible to the stochastic effects. Furthermore, the evolution through the early part of the RSG branch is very fast. In light of this, we correlate \lmin\ with cluster age in a probabilistic sense using population synthesis. 

\begin{figure}
\caption{Plot showing the most likely RSG \lmin\ for a cluster at a given age containing 50 RSGs, from MIST non-rotating isochrones \citep{dotter2016mesa}. Each point represents the median \lmin\ of 1000 trials while the error bars represent the 68\% probability limits. For clusters with fewer RSGs the relation stays the same but there are more stochastic errors. }
\centering 
\label{fig:lminpdf}
\includegraphics[width=\columnwidth]{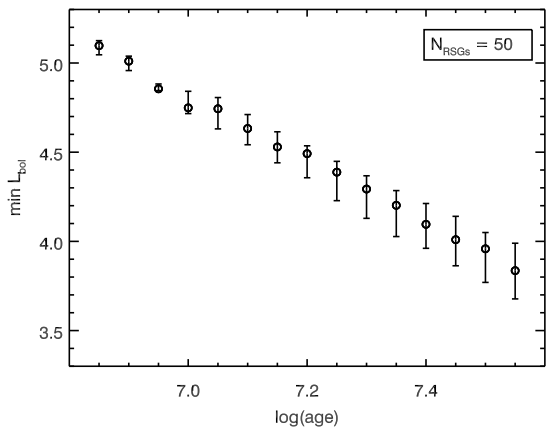}
\end{figure}

We again utilise the MIST grid of isochrones. At each age, we identify the RSG phase as being where \teff $\leq$ 4500K and \logl\ $\geq$ 4. Next, we identify the minimum and maximum stellar mass from the defined RSG phase and generate a sample of 500 synthetic stars following a Salpeter IMF, and then take a random subsample of these stars to match the number of real RSGs in a given cluster, e.g. for NGC 2100 there are 19 RSGs. At each trial, we determine the value of \lmin\ and repeat 1000 times to get a probability density function. Figure \ref{fig:lminpdf} shows the most likely \lmin\ values for a cluster containing 50 RSGs at each age in our grid. This process is repeated at each age, yielding a 2-D probability density function of age and \lmin. From this we interpolate the observed \lmin\ for a cluster, which itself is randomly sampled 1000 times from a Gaussian distribution to take into account the errors, and derive an age distribution for the cluster. Results are shown in Table 2. As the errors on \lbol\ are small the error on age is taken from the nearest grid point.

\begin{figure*}
\caption{Plot showing best fitting isochrones from each of the three age estimators for the Galactic clusters. The solid portion of the isochrones represents the main sequence. The purple star symbol indicates the brightest TO star used for the age determination. }
\centering
\label{fig:Galiso}
\includegraphics[width=\columnwidth]{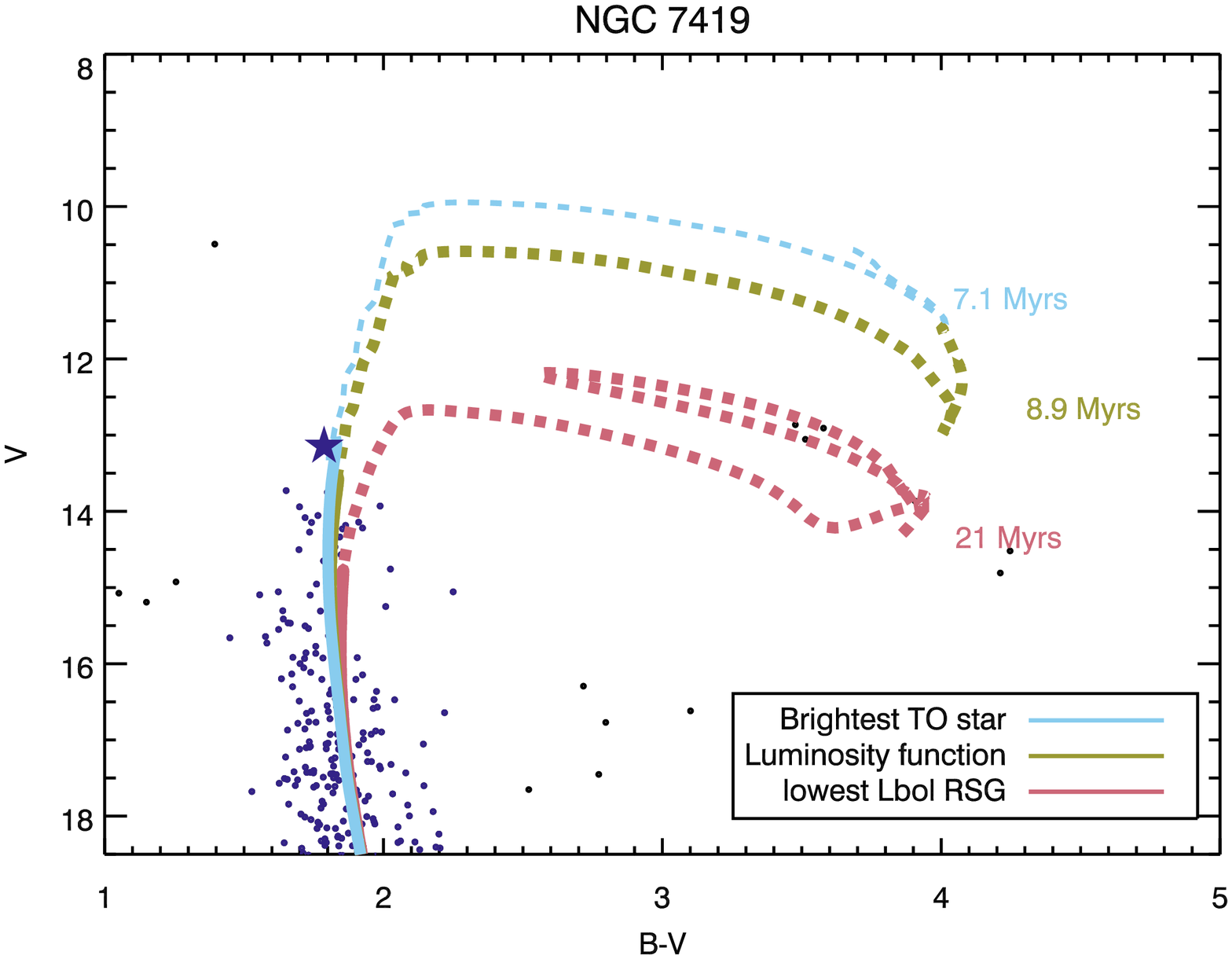}\includegraphics[width=\columnwidth]{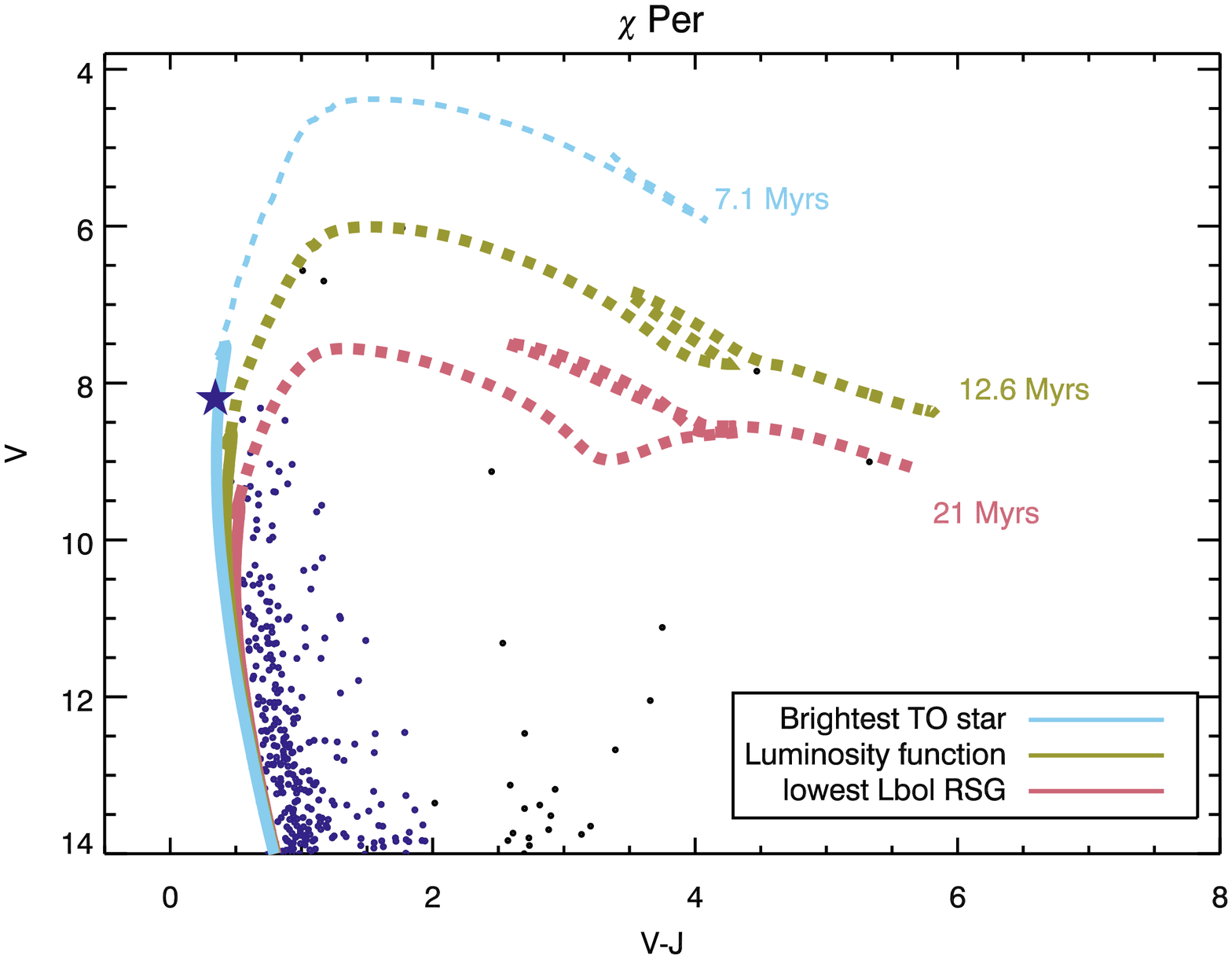}
\includegraphics[width=\columnwidth]{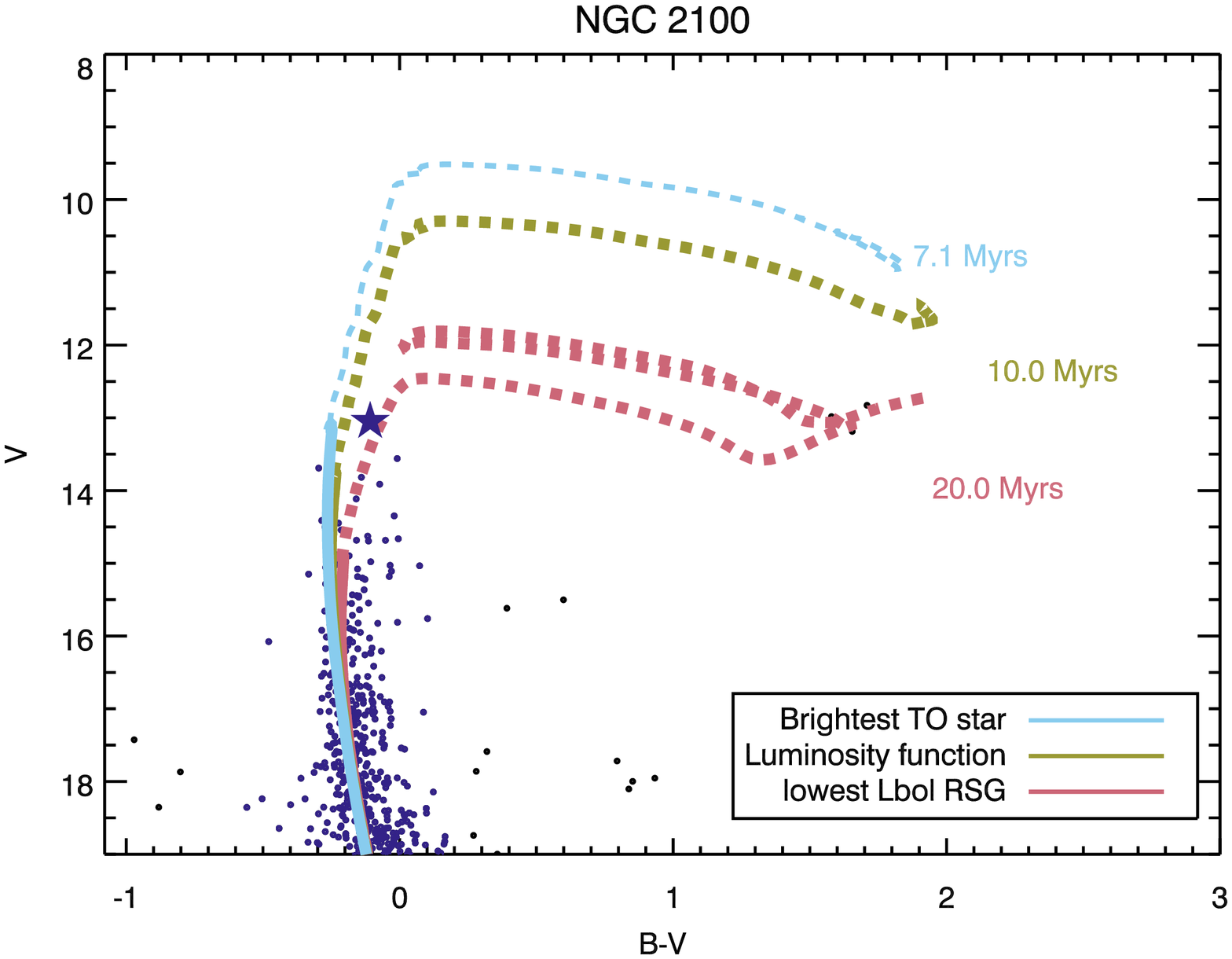}\includegraphics[width=\columnwidth]{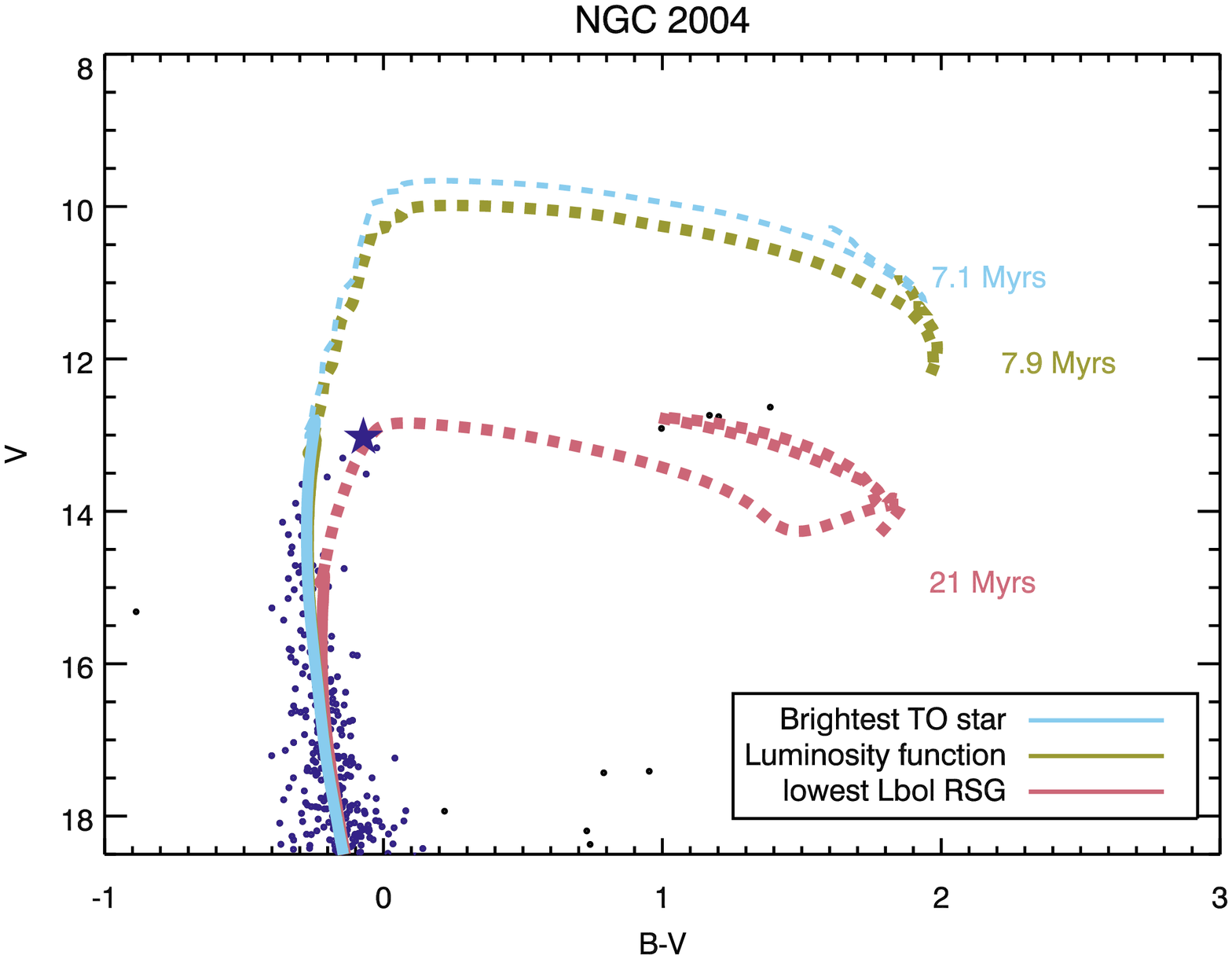}
\end{figure*}

\section{Results}
The age estimates for each cluster are listed in Table 2. Figure \ref{fig:Galiso} shows the best fitting isochrones determined for each cluster via each of the methods described above. Note that often the RSGs were saturated in these filters and hence could not be shown. Due to the uncertainty in the colours of late-time stars the isochrones often do not sit where RSGs are observed. In these plots the red line shows the results from the lowest luminosity RSG method, the green line shows the result from fitting the luminosity function of the TO and the blue line shows the results from fitting the single brightest star at the MSTO. The solid portion of the line shows the MS. Each isochrone has been reddened using the extinctions determined in Section 3.1 and the extinction coefficients from \citet{cardelli1988ext} (optical) and \citet{koornneef1983near} (mid-IR).
\begin{table*}

\centering

\caption{Age estimations for each cluster found using the three different age determination methods. For the brightest TO method and the lowest \lbol\ RSG method, the errors on photometry were small and hence the errors presented here represent the grid spacing of of the ages. }
\label{results}
\begin{tabular}{lccccccccc}

\hline
Cluster & \multicolumn{9}{|c|}{Age estimates (Myr)} \\[0.5ex]
 & \multicolumn{3}{|c|}{Brightest TO star} &\multicolumn{2}{|c|}{Luminosity function}  &\multicolumn{4}{|c|}{Lowest \lbol\ RSG}    \\ 
&Magnitude &Rotating & Non-rotating & Rotating & Non-rotating &N${_{\rm RSGs}}$&\lbol\ &Rotating & Non-rotating \\
\hline
NGC 7419 &13.14 (V)& 8.3 $\pm$ 1 & 7.1 $\pm$ 1 & 9$^{+1}_{-1}$ & 9$^{+2}_{-1}$  &5& 4.37&21 $\pm$ 1 & 20$\pm$ 1   \\
$\chi$ Per &8.19 (V) & 8.1 $\pm$ 1 & 7.9 $\pm$ 1 & 13$^{+2}_{-2}$  & 13$^{+2}_{-2}$  &8 &4.38&22 $\pm$ 1 & 21 $\pm$ 1    \\
NGC 2100 &13.05 (V)&7.1 $\pm$ 1 & 7.1 $\pm$ 1 & 10$^{+1}_{-1}$  & 10$^{+1}_{-1}$  &19&4.43& 22 $\pm$ 1  & 20 $\pm$ 1    \\
NGC 2004 &13.02 (V) &7.1 $\pm$ 1 &6.3 $\pm$ 1 & 8$^{+1}_{-0.5}$  & 8$^{+1}_{-0.5}$  &7& 4.35&24 $\pm$ 1   & 22 $\pm$ 1   \\

\hline

\end{tabular}
\end{table*}

For all of the clusters, the difference between the ages determined using the brightest TO star and the luminosity function of the cluster TO is minimal and smaller than the difference between rotating and non-rotating models, at most the difference in best fit age varies by 0.5Myr, but they are all consistent to within the errors. The lowest luminosity RSG method always finds the cluster to be older than the cluster TO methods. The most extreme example of this is NGC 2004, for which the RSG method finds an age that is three times older. In the following sections we will discuss the results for each cluster in turn and compare to previous age estimates. 

\subsection{NGC 7419}
This cluster has previously been studied by \citet{marco2013ngc}, who fit isochrones from \citet{marigo2008evolution} by eye to UBV data of the entire cluster, finding an age of 14Myr. The authors found that a 12Myr isochrone was a good fit to the TO once B stars were removed, but an older age of 15Myr also fit the apparent TO well while leaving some brighter stars above. \citet{beasor2018evolution} found the previously best fitting age isochrone (14 Myr, Padova) found by \citep{marco2013ngc} was equivalent to an age of 10-14 Myr if the Geneva isochrones were used to fit the same MSTO point.  

In this study, the age estimates for NGC 7419 vary between methods by approximately 10 Myr for both rotating and non-rotating models. As with all the clusters, the brightest TO and the luminosity function of the TO methods (~7-8 Myr) suggest the cluster is far younger than the age determined using the lowest \lbol\ RSG ($\sim$17 Myr). There is very little difference between the ages found for the cluster using the brightest TO star and the luminosity function of the TO. The previous age estimates for this cluster are approximately an average of the TO and RSG ages we present here, as previous studies have fit the TO and RSG branch diagnostics simultaneously.

\subsection{$\chi$ Persei}
This cluster is part of the $h$ and $\chi$ Per Double Cluster, and has been studied before in detail by \citet{currie2010stellar}. Using UBVI data, the age of the cluster was estimated by these authors a number of ways. Using the MS they find an age of 14 Myrs, and claim the M supergiants support this age. Using Padova isochrones they fit by eye to see which best fits the RSG range, however they use the full luminosity range to make a prediction, and as previously discussed due to evolution and the possibility of post-merger objects there is no unique relation for age and the brightest RSGs. In \citet{beasor2018evolution} the best fitting Padova isochrone was again used to find a model dependent age for the cluster, finding again a Geneva age estimate of 10-14 Myr (for rotating and non-rotating models).

In this work we find ages ranging from 8 Myr to 18 Myr (with the TO method giving the youngest age estimate and the RSG method giving the oldest). Again the previously presented age estimates are somewhere in the middle of this. While \citet{currie2010stellar} did not fit the entire cluster (TO and M-supergiants) simultaneously we find the age given by the luminosity function method gives an age consistent with theirs. The cause for discrepancy when using the RSGs between \citeauthor{currie2010stellar} and our work could be simply due to the previous authors using the entire luminosity range of the RSGs to estimate an age, rather than the lowest \lbol RSG.


\subsection{NGC 2100}
There are a number of age estimations for LMC cluster NGC 2100 in the literature. \citet{niederhofer2015no} fit both the TO and the RSG branch simultaneously, giving an age of 21 Myr, in agreement with the age found from our lowest \lbol\ RSG method. We again see that at this age the model Hess diagram cannot explain the brightest blue stars above the MSTO, and the authors claim this is an effect likely caused by stellar rotation (see Section 5.1).  The age has been estimated since then by \citet{beasor2016evolution} who used the full range of RSG luminosities and evolutionary mass tracks (see Fig. 8 within the paper).  This gave a younger age of $\sim$15 Myr. 

In this work we find an age  discrepancy of a factor of 2 between the brightest TO star method and the RSG method. We find using the MSTO gives an age of 7-10 Myr (for the single brightest TO star and LF respectively). For the lowest \lbol\ RSG method we find an age of 20-22Myr, older than found by \citet{beasor2016evolution} by 5 Myr. 
\subsection{NGC 2004}
This LMC cluster was studied in \citet{niederhofer2015no}, in which the authors estimated the cluster age to be 20 Myr, but again the observed MS reached brighter magnitudes than predicted by the model. This age is also consistent with the age derived from our lowest \lbol\ RSG method, likely because the authors disfavour using the brightest stars in the TO as they believe them to be either BSSs or evidence for an extended main sequence TO (see Section 5.1). This cluster has the largest discrepancy in age estimations out of all the clusters in this sample, with the RSG method suggesting an age that is up to 4 times older than the TO method results.
\section{Discussion}
The analysis has shown that age estimates vary by up to a factor of 4 depending on which method is employed, see Fig. \ref{fig:Galiso}. The most dramatic difference is seen for NGC 2004, where the brightest TO star gives an age of 6 and 7 Myr for non-rotating and rotating models respectively, while the RSG method gives an age of 22 Myr or 24 Myr  for non-rotating and rotating models respectively. We will now discuss possible causes for these age discrepancies in detail. 
\subsection{Possible causes for age discrepancies}
Stellar rotation causes extra hydrogen to be mixed into the core of the star, prolonging the life of a star. Stars that rotate more rapidly spend longer on the MS than stars that rotate more slowly, making it difficult to explain all of the stars at the TO with a single age isochrone. 

For young and intermediate age clusters \citep[20Myr - $\sim$2Gyr,][]{bastian2009effect}, rotation has been used to explain the extended main sequence turn off (eMSTO) phenomenon, where the TO of a cluster appears brighter than expected in a colour magnitude diagram \citep[e.g.][]{keller2011extended}.  In \citet{niederhofer2015no}, rotating and non-rotating isochrones were used to determine what perceived age spread would be seen in a cluster due to stellar rotation. This study showed that as clusters increase in age, the apparent age spread due to the eMSTO increases (i.e. clusters with ages <100 Myr have small age spreads, on the order of tens of Myr, while clusters $\sim$ 1 Gyr have apparent age spreads on the order of a few hundred Myr). This age spread is proportional to the age of the isochrone. While this study implies the eMSTO effect is likely to be present in the clusters presented in this work, isochrones used in \citet{niederhofer2015apparent} were between 50 Myr - 1 Gyr, older than the 4 clusters presented in our study. Therefore the significance of the eMSTO at ages < 50 Myr is not yet quantified.  

To check whether rotation could have a significant effect on the ages determined for the clusters in our sample, we first used MIST rotating and non-rotating models for all ages; Table 1 shows the results for all methods. The rotating MIST models include stars with rotation at 0.4 of the critical velocity. We can see from the results that the difference in ages between rotating and non-rotating models is minimal ($\pm$10\%). Therefore, stars rotating at this speed are not able to explain the differences in ages we obtain between the methods utilising the MSTO of the cluster and the method using the red stars. Therefore, to extend the MS by long enough to explain the age differences, stars would have to be rotating at speeds faster than 0.4 of the critical velocity. 

\begin{figure}
\caption{Plot showing the CMD for the stars in NGC 2100. Overplotted are isochrones from \citep{georgy2013grids} (the rotating and non-rotating 20 Myr, age from RSG method) and MIST (10 Myr, age from the luminosity function method). Filled coloured circles represent the MSTO for each isochrone. }
\centering
\label{fig:genevarot}
\includegraphics[width=\columnwidth]{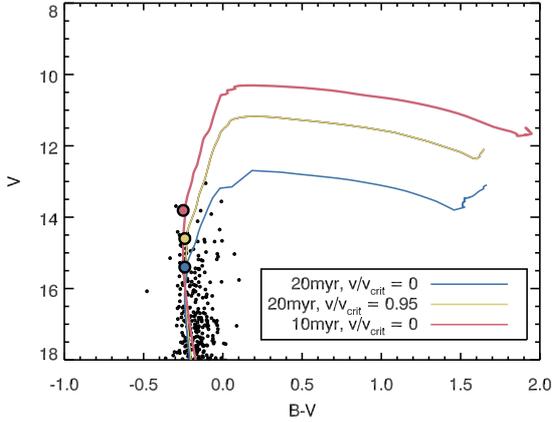}

\end{figure}

To further investigate the impact of stellar rotation on inferred ages, we have re-analysed our data for NGC 2100 with the isochrones of \citet{georgy2013grids}, which have a much sparser sampling in stellar mass but which explore a much greater range of initial rotation rates. Figure \ref{fig:genevarot} shows the colour magnitude diagram for NGC 2100 with 3 isochrones overplotted. The first is the 10 Myr MESA isochrone, which we found to be the best fitting age when using the LF method, shown by the red line. The second is the 20Myr non-rotating isochrone from \citeauthor{georgy2013grids}, the best fitting age found when using the lowest \lbol\ RSG, shown by the blue line. Finally, we also show the 10 Myr iscochrone where the stars are rotating at 0.95 of the critical velocity, shown by the yellow line. We have highlighted the tip of the TO according to each model with a circle.

Figure \ref{fig:genevarot} shows the TO of the best fitting isochrone found from the LF method is 1.3 mag brighter than the TO of the best fitting isochrone from the RSG method, when there is no rotation. When stellar rotation is present, even with stars rotating at 0.95 of critical the difference in magnitude between the 10 Myr TO and the 20 Myr TO is 0.8 mag.  These results argue that even extreme rotation rates cannot explain the age discrepancies we find. Further to this, \citet{marco2013ngc} counted the number of Be stars, objects showing H$\alpha$ emission commonly thought to be fast rotators, within the MS of NGC 7419 and find the number of Be stars at the TO is approximately equal to the number of `normal' objects (see Fig. 8 within their paper). Therefore, while stellar rotation is likely to have some effect on broadening the MSTO, it cannot fully account for the discrepancies we have observed. We speculate that stars above the TO of the fast rotating isochrone may be BS-like stars formed via binary interaction, i.e. merger products or mass gainers. 

We now compare our results to the predictions of \citet{schneider2015evolution}, who computed the evolution of single and close interacting binary stars. They predict the ratio of the number of blue stragglers expected in a cluster ($N_{\rm bss}$), to the number of stars 2 magnitudes below the TO ($N_{\rm 2}$) as a function of cluster age (see Fig. 14 within that paper). When taking the same diagnostics for NGC 2100  (using the TO from the Geneva rotating models as shown by the yellow dot in Fig. \ref{fig:genevarot}) we find an $N_{\rm bss}$/$N_{\rm 2}$ ratio of 0.11, in close agreement with the predictions of \citeauthor{schneider2015evolution}, who predict a ratio of 0.1-0.2 for a 20Myr cluster. All of the potential BS-star candidates\footnote{The exact number of BS candidates changes slightly depending on which isochrones are used. Here we have chosen to use the Geneva isochrones rotating at 0.95 of the critical velocity to see a lower limit on the number of BS candidates.} in NGC 2100 were found to be cluster members by \citet{niederhofer2015no}. Further to this, Figure 16 in \citet{schneider2015evolution} shows the apparent stellar age, $\tau_*$, of the most massive blue straggler star as a function of the true cluster age, $\tau_{\rm cl}$. For a 20Myr cluster, the most massive BSS would cause an inferred age of approximately 5-10Myr, again in good agreement with our findings. It should be noted however that the predictions of \citeauthor{schneider2015evolution} are based on theoretical clusters with a primordial binary fraction of 100\%, so the age discrepancies and number of BSSs they present are upper limits. However, it is encouraging to see that these model predictions are comparable to our observations.

\subsection{How should we determine ages for young clusters?}
Initial masses for post-MS transitional objects such as luminous blue variables (LBVs), Wolf-Rayet stars and RSGs are often estimated by locating such objects in clusters, then determining the cluster age using the cluster TO and single rotation isochrones \cite[e.g.][]{humphreys1985on,massey2000progenitor, massey2001progenitor}. Throughout this work we have shown that using the TO is likely to cause the age inferred to be too young, and hence the initial masses too high. We attribute this effect to BS-like stars above the TO, some of which may be explained by a distribution of rotation rates at ZAMS. To attempt to quantify this effect, and to determine which age estimation method provides the most accurate results, we repeat our analysis on synthetic clusters of a known age with a realistic rotation distribution, described below. 

We now enlist synthetic clusters from Geneva \citep{ekstrom2012grids} to investigate the impact of using the cluster TO more thoroughly. We generated a 20 Myr old cluster at Solar metallicity (Z=0.014), where the stellar rotations follow the distribution of \citet{huang2010stellar}. Next, the mass of the cluster is randomised between 0.5-5$\times$10$^4$M$_\odot$, typical of clusters known to contain RSGs. From this subsample, we can identify the brightest TO star and the least luminous RSG (where the RSG phase is defined as where \teff\ $\leq$ 4500K). To ensure we are comparing like-with-like, we use Geneva non-rotating isochrones to determine age estimations via the methods described in Section 3.2 and 3.4. Our grid of Geneva models has ages from 6 to 25 Myrs. We repeat this for 1000 trials, where each trial generates a new cluster with varying cluster mass, thus allowing us to see how the age discrepancy between each method varies with total cluster mass (M$_{\rm tot}$).  

\begin{figure}
\caption{Ages derived from age fitting a 20Myr synthetic cluster containing only single stars using the TO method and the lowest \lbol\ RSG method. The dashed line shows the true age of the cluster. }
\centering
\label{fig:fakeclust}
\includegraphics[width=\columnwidth]{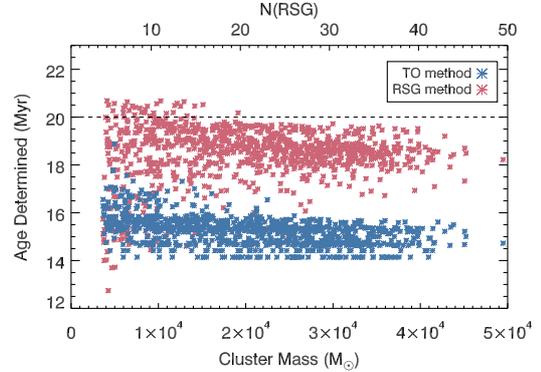}
\end{figure}

Figure \ref{fig:fakeclust} shows the distribution of age estimations for each method as a function of cluster mass. The dashed line shows the true cluster age and the results of each trial are plotted with blue crosses (TO method) or red crosses (RSG method). For the RSG method in particular, the errors on age are larger at lower cluster masses. This is due to lower mass clusters perhaps only containing $\sim$1 RSG, leading to large stochastic errors. 

Our results demonstrate clearly that using the cluster TO method can cause a systematic underestimation in age by $\sim$25\% due to rotation on the MS extending the lifetime of MS stars and causing them to appear more luminous. While there is also a systematic offset when using the lowest \lbol\ RSG method, this is a much smaller effect, on the order of $\sim$10\%. This offset is due to numerical effects when interpolating mass tracks containing blue loops, causing the bottom of the RSG branch to appear more luminous. For a typical cluster where M$_{\rm tot}$ < 5$\times$10$^{4}$M$_\odot$ the random error is approximately 5-10\%, comparable to the systematic offset. 

\begin{figure}
\caption{Ages derived from age fitting a 20Myr synthetic cluster using the TO method and the lowest \lbol\ RSG method. In this case, the synthetic cluster has a binary fraction of 50\%. The dashed line shows the true age of the cluster. }
\centering
\label{fig:fakeclustbin}
\includegraphics[width=\columnwidth]{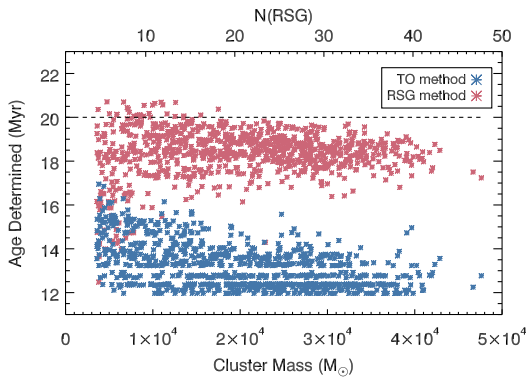}
\end{figure}

 We now repeat this experiment and include unresolved binaries with a binary fraction of 50\%\footnote{As the isochrones used to create each cluster rely on single star tracks it is not possible to account for the evolution of interacting binary systems \citep{georgy2014syclist}}. Figure \ref{fig:fakeclustbin} shows the determined ages for a cluster as a function of cluster mass with a binary fraction of 50\%, as shown by the blue crosses. This illustrates that the inclusion of unresolved binaries further reduces the usefulness of the TO method, causing an even greater discrepancy ($\sim$ 6Myr), while the RSG method remains much less affected (note that the red crosses on the plot show the results for the RSG method from both the cluster containing only single stars, and the cluster with a binary fraction of 50\%). This is because the RSGs are significantly more luminous than the companion stars and hence are unaffected by unresolved binaries. From our analysis we can conclude that using the lowest \lbol\ RSG method is the most reliable. Interestingly, the age discrepancies found by this experiment are far lower than the discrepancies observed in the real clusters. This supports our conclusions that the age discrepancies we observe for the 4 real clusters cannot be caused by rotation on the MS alone, and are likely caused by a combination of rotation, unresolved binaries and binary products (such as mass gainers or mergers).

\section{Conclusions}
In this work we present age estimates for 2 Galactic clusters and 2 LMC clusters using 3 different methods. By doing this we have demonstrated a systematic offset between ages determined using the MS stars in a cluster and using the evolved RSGs.  Our results cannot be explained by rotation on the MS alone, and instead we suggest that the age discrepancies are caused by a combination of rotation, unresolved binaries and binary products (e.g. mergers and mass gainers). We also demonstrate using synthetic clusters that using the lowest \lbol\ RSG method will yield the most reliable age estimation. 

\section*{Acknowledgements}
The authors would like to thank the anonymous referee for interesting comments which helped improve the paper. E.R.B. acknowledges support from an STFC doctoral studentship. N.B. gratefully acknowledges financial support from the Royal Society (University Research Fellowship) and the European Research Council (ERC-CoG-646928-Multi-Pop). This work makes use of {\tt IDL} software packages and astrolib. 




\bibliographystyle{mnras}
\bibliography{references} 




\appendix


\bsp	
\label{lastpage}
\end{document}